\documentclass{article} 
\usepackage{nips12submit_e,times}

\title{ Affinity Weighted Embedding }

\author{
Jason Weston \\
Google Inc.,\\
New York, NY, USA. \\
\texttt{jweston@google.com} \\
\And
Ron Weiss \\
Google Inc.,\\
New York, NY, USA. \\
\texttt{ronw@google.com} \\
\AND
Hector Yee \\
Google Inc.,\\
San Bruno, CA, USA. \\
\texttt{hyee@google.com} \\
}

\nipsfinalcopy 

\begin{document}

\maketitle

\begin{abstract}
Supervised (linear) embedding models like Wsabie \cite{weston:2011:ijcai} and
PSI \cite{psi} have proven successful at 
ranking, recommendation and annotation tasks.
However, despite being scalable to large datasets they do not take
full advantage of the extra data due to their linear nature,
and typically underfit.
We propose a new class of models which aim to provide improved 
performance while retaining many of the benefits of the existing class
of embedding models.
Our new approach 
works by iteratively
learning a linear embedding model where the next iteration's
features and labels are reweighted as a function of 
the previous iteration. 
We describe several variants of the family, and give 
some initial results.
\end{abstract}

%

\section{(Supervised) Linear Embedding Models}

Standard linear embedding models are of the form:
\[
  f(x, y) = x^\top U^\top V y  =  \sum_{ij} x_i U_i^\top V_j y_j.
\]
where $x$ are the input features and $y$ is a
possible label (in the annotation case), document (in the information retrieval case)
or item (in the recommendation case).
These models are used in both supervised and unsupervised settings.
In the supervised ranking case, they have proved successful in many of the tasks
described above, e.g. the Wsabie algorithm 
\cite{weston:2011:ijcai,music-wsabie,weston2012latent} which  approximately optimizes precision
at the top of the ranked list has proven useful for annotation and recommendation.

These methods scale well to large data and are simple to implement and use.
However, as they contain no nonlinearities (other than in the feature representation
in $x$ and $y$) they can be limited in their ability to fit large complex datasets,
and in our experience typically underfit.

\section{Affinity Weighted Embedding Models}

In this work we propose the following generalized  embedding model:
\[
  f(x, y) = \sum_{ij} G_{ij}(x,y) ~ x_i U_i^\top V_j y_j.
\]
where $G$ is a function, built from a previous learning step,
that measures the affinity between two points. 
Given a pair $x$, $y$ and feature indices $i$ and $j$, $G$ returns a scalar.
Large values of the scalar indicate a high degree of match.
Different methods of learning (or choosing) $G$ lead to different variants of our
proposed approach:
\begin{itemize}
\item $G_{ij}(x,y) = G(x,y)$. In this case each feature index pair $i$, $j$ returns the same scalar
so the model reduces to:
\[
  f(x, y) =  G(x,y)  ~ x^\top U^\top V y.
\]
\item $G_{ij}(x,y) = G_{ij}$. In this case the returned scalar for $i$, $j$ is the same independent
of the input vector $x$ and label $y$, i.e. it is a reweighting of the feature pairs.
This gives the model:
\[
  f(x, y) =  \sum_{ij} G_{ij} x_i U_i^\top V_j y_j.
\]
This is likely only useful in large sparse feature spaces, e.g. if $G_{ij}$ represents the 
weight of a word-pair in an information retrieval task or an item-pair in a recommendation task.
Further, it is possible that $G_{ij}$ could take a particular form, e.g. it is represented as
a low rank matrix  $G_{ij} = g_i^\top g_j $.
In that case we have the model $   f(x, y) =  \sum_{ij} g_i^\top g_j x_i U^\top_i V_j y_j$.
\end{itemize}

While it may be possible to learn the parameters of $G$ jointly with $U$ and $V$ here
we advocate an iterative approach:
\begin{enumerate}

\item Train a standard embedding model: $ f(x, y) = x^\top U^\top V y$.

\item Build $G$ using the representation learnt in (1).

\item Train a weighted model: $ f(x, y) = \sum_{ij} G_{ij}(x,y) ~ x_i \bar{U}_i^\top \bar{V}_j y_j$.

\item Possibly repeat the procedure further: build $\bar{G}$ from (3).  (So far we have not tried this).
\end{enumerate}

Note that the training algorithm used for (3) is the same as for (1) -- we only change the model.

In the following,  we will focus on the $G_{ij}(x,y) = G(x,y)$ case
(where we only weight examples, not features) 
and a particular choice of $G$\footnote{Although perhaps 
$ G(x,y) =  \sum_{i=1}^m \exp(-\lambda_x ||U x- U {\bf{x}}_i||^2) \exp(-\lambda_y ||V y- V {\bf{y}}_i||^2)$
 would be more natural. Further we could also consider 
   $  G_{orig}(x,y) =  \sum_{i=1}^m \exp(-\lambda_x ||x- {\bf{x}}_i||^2) \exp(-\lambda_y ||y- {\bf{y}}_i||^2)$
which does not make use of the embedding in step (1) at all. This would likely perform poorly
when the input features are too sparse, which would be the point of improving the representation
by learning it with $U$ and $V$.}:
\begin{equation} \label{k-eq}
     G(x,y) =  \sum_{i=1}^m \exp(-\lambda_x ||U x- U {\bf{x}}_i||^2) \exp(-\lambda_y || y- {\bf{y_i}}||^2)
\end{equation}
where ${\bf{x}}$ and ${\bf{y}}$ are the sets of vectors from the training set.

$G$ is built using the embedding $U$ 
learnt in step (1), and is then used to build a new embedding model in step (3). 
Due to the iterative nature of the steps we can compute $G$ for all examples
in parallel using a MapReduce framework,
and store the training set necessary for step (3), thus making learning straight-forward.
To decrease storage, instead of computing a smooth $G$ as above we can clip (sparsify) $G$
by taking only the top $n$ nearest neighbors to $U x$, and set the rest to 0.
Further we take $\lambda_y$ suitably large such that $\exp(-\lambda_y || y- {\bf{y_i}}||^2)$ either gives
1 for ${\bf{y}}_i = y$  or 0 otherwise\footnote{This is useful in the label annotation or item ranking settings, but would not be a good idea in an information retrieval setting.}.
In summary, then, for each training example,
we simply have to find the ($n=20$ in our experiments) nearest neighboring examples in the embedding space,
and then we reweight their labels using eq. \ref{k-eq}. (All other labels would then receive a weight of zero,
although one could also add a constant bias to guarantee those labels can receive non-zero final scores.)

\section{Experiments}

So far, we have conducted two preliminary experiments on Magnatagatune (annotating music with text tags) 
and ImageNet (annotation images with labels).
Wsabie has been applied to both tasks previously \cite{music-wsabie,weston:2011:ijcai}.

\begin{table}[h]
\caption{Magnatagatune Results}
\label{tagatune-results}
\begin{center}
\begin{tabular}{llll}
\multicolumn{1}{c}{\bf Algorithm}  
&\multicolumn{1}{c}{\bf Prec@1}
&\multicolumn{1}{c}{\bf Prec@3}
\\ \hline \\
$k$-Nearest Neighbor                   & 39.4\%   & 28.6\% \\
$k$-Nearest Neighbor (Wsabie space)    & 45.2\%   & 31.9\% \\
Wsabie                                 & 48.7\%   & 37.5\% \\
Affinity Weighted Embedding
                                       & 52.7\%   & 39.2\% \\
\end{tabular}
\end{center}
\end{table}

\begin{table}[h]
\caption{ImageNet Results (Fall 2011, 21k labels)}
\label{imagenet-results}
\begin{center}
\begin{tabular}{ll}
\multicolumn{1}{c}{\bf Algorithm}  &\multicolumn{1}{l}{\bf Prec@1}
\\ \hline \\
Wsabie (KPCA features)                  & ~9.2\% \\
$k$-Nearest Neighbor (Wsabie space)         & 13.7\% \\
Affinity Weighted Embedding                                     & 16.4\% \\
Convolutional Net \cite{dean2012large}  & 15.6\% {\tiny (NOTE: on a different train/test split)} \\
\end{tabular}
\end{center}
\end{table}

On Magnatagatune we used MFCC features for both Wsabie and our method, similar to those used  in \cite{music-wsabie}.
For both models we used an embedding dimension of 100.
Our method improved over Wsabie marginally as shown in Table \ref{tagatune-results}.
We speculate that this improvement is small due to the small size of the dataset
(only 16,000 training examples, 104 input dimensions for the MFCCs and 160 unique tags).
We believe our method will be more useful on larger tasks.

On the ImageNet task (Fall 2011, 10M examples, 474 KPCA features and 21k classes)
 the improvement over Wsabie is much larger, shown in Table \ref{imagenet-results}.
We used similar KPCA features as in \cite{weston:2011:ijcai} for both Wsabie and our method. We
use an embedding dimension of 128 for both.
We also compare to nearest neighbor in the embedding space.
For our method, we used the max instead of the sum in eq. (\ref{k-eq}) as it gave better results.
Our method is competitive with the convolutional neural network model of \cite{dean2012large} (note, this is on a different train/test split).
However, we believe the method of \cite{krizhevsky2012imagenet}
 would likely perform better again if applied in the same setting.

\section{Conclusions}

In conclusion, by incorporating a learnt reweighting function $G$ 
into supervised linear embedding we can increase
the capacity of the model leading to improved results.
One issue however is that the cost of reducing underfitting by using $G$
is that it both increases the storage and computational requirements of the model. One avenue we have begun 
exploring in that regard is to use approximate methods in order to compute $G$.


\small
\bibliography{biblio}

\begin{thebibliography}{1}

\bibitem{psi}
B.~Bai, J.~Weston, D.~Grangier, R.~Collobert, K.~Sadamasa, Y.~Qi, C.~Cortes,
  and M.~Mohri.
\newblock Polynomial semantic indexing.
\newblock In {\em NIPS}, 2009.

\bibitem{dean2012large}
J.~Dean, G.~Corrado, R.~Monga, K.~Chen, M.~Devin, Q.~Le, M.~Mao, A.~Senior,
  P.~Tucker, K.~Yang, et~al.
\newblock Large scale distributed deep networks.
\newblock In {\em Advances in Neural Information Processing Systems 25}, pages
  1232--1240, 2012.

\bibitem{krizhevsky2012imagenet}
A.~Krizhevsky, I.~Sutskever, and G.~Hinton.
\newblock Imagenet classification with deep convolutional neural networks.
\newblock In {\em Advances in Neural Information Processing Systems 25}, pages
  1106--1114, 2012.

\bibitem{music-wsabie}
J.~Weston, S.~Bengio, and P.~Hamel.
\newblock Large-scale music annotation and retrieval: Learning to rank in joint
  semantic spaces.
\newblock In {\em Journal of New Music Research}, 2012.

\bibitem{weston:2011:ijcai}
J.~Weston, S.~Bengio, and N.~Usunier.
\newblock Wsabie: Scaling up to large vocabulary image annotation.
\newblock In {\em Intl. Joint Conf. Artificial Intelligence, ({IJCAI})}, pages
  2764--2770, 2011.

\bibitem{weston2012latent}
J.~Weston, C.~Wang, R.~Weiss, and A.~Berenzeig.
\newblock Latent collaborative retrieval.
\newblock {\em ICML}, 2012.

\end{thebibliography}
\bibliographystyle{plain}

\end{document}